\documentclass[12pt]{article} 

\usepackage{epsfig}
\usepackage{graphicx}
\usepackage{amsmath}
\usepackage{amssymb}
\usepackage{bm}

\usepackage[ps2pdf,%
a4paper=true,%
breaklinks=true,%
colorlinks=true,%
pdfauthor={First Author et al.},%
pdftitle={Template for manuscripts in Advances in Space Research}%
]{hyperref}
\usepackage{url}
\usepackage{breakurl}

\begin{document}

\title{A possible flyby anomaly for Juno at Jupiter}


\author{L. Acedo\thanks{E-mail: luiacrod@imm.upv.es}, P. Piqueras and J. A. Mora\~no \\
Instituto Universitario de Matem\'atica Multidisciplinar,\\
Building 8G, $2^{\mathrm{o}}$ Floor, Camino de Vera,\\
Universitat Polit$\grave{\mbox{e}}$cnica de Val$\grave{\mbox{e}}$ncia,\\
Valencia, Spain\\
}

\maketitle

\begin{abstract}
In the last decades there have been an increasing interest in improving the accuracy of spacecraft
navigation and trajectory data. In the course of this plan some anomalies have been found
that cannot, in principle, be explained in the context of the most accurate orbital models including
all known effects from classical dynamics and general relativity.
Of particular interest for its puzzling nature, and the lack of any accepted explanation for the moment,
is the flyby anomaly discovered in some spacecraft flybys of the Earth over the course of twenty years.
This anomaly manifest itself as the impossibility of matching the pre and post-encounter Doppler tracking and
ranging data within a single orbit but, on the contrary, a difference of a few mm$/$s in the asymptotic
velocities is required to perform the fitting.

Nevertheless, no dedicated missions have been carried out to elucidate the origin of this phenomenon with
the objective either of revising our understanding of gravity or to improve the accuracy of spacecraft
Doppler tracking by revealing a conventional origin.

With the occasion of the Juno mission arrival at Jupiter and the close flybys of this planet, that are currently
been performed, we have developed an orbital model suited to the time window close to the perijove. This model
shows that an anomalous acceleration of a few mm$/$s$^2$ is also present in this case. The chance for overlooked
conventional or possible unconventional explanations is discussed.
\end{abstract}


\section{Introduction}
\label{intro}
A key step towards interplanetary space exploration was achieved by the theoretical work of Minovitch \cite{Minovitch1,Minovitch2} and Flandro \cite{Flandro}.
In the early sixties of the past century these authors proposed the use of the gravitational assist manoeuvre to increase
the energy of spacecraft in the Solar System barycenter, allowing for fast reconnaissance missions to the outer planets from
Jupiter to Neptune \cite{Butrica}. Since then, many gravity assist, flyby or slingshot manoeuvers (as this manoeuver can be equally be named) have been programmed in the course of missions to the inner planets (Mariner, Messenger),
outer planets (Pioneer, Voyager, Galileo, Cassini, New Horizons, Juno) or asteroids (NEAR). The objective of many of these flybys
is to obtain data from the planets as they flyby them and to take advantage of the energy transfer obtained during the
flyby \cite{Transfer}.

Apart from the obvious contribution to planetary science, these missions have provided an excellent framework to perform
tests of General Relativity and to improve the accuracy of trajectory determination systems. As soon as 1976, the Viking 
mission allowed for the verification of Shapiro's echo delay prediction of an increase in a time taken for a round-trip's light 
signal to travel between the Earth and Mars as a consequence of the curvature of space-time by the Sun \cite{VikingExp}. More recently, Everitt et al. \cite{Everitt} have tested the geodetic and frame-dragging effects. Also, the analysis of the data from the Messenger mission to Mercury is now used for improving the accuracy of ephemeris as they also put a stringent test on
the parameters in the post-newtonian formalism \cite{Messenger1,Messenger2}. With such an ongoing interest in fundamental aspects
of spacecraft dynamics and gravity it is, perhaps, not surprising that some anomalies have showed up in the years passed since the beginning of the space age. Among them, the so-called Pioneer anomaly stands out as a particularly interesting case. As it has become common lore within the space physics community, the Pioneer anomaly consist on a trend detected on the Doppler 
data for the Pioneer 10 and Pioneer 11 spacecraft as they travel beyond Jupiter. This trend was consistent with an, almost
constant, acceleration of $a_P=(8.74\pm 1.33) \times 10^{-8} $ cm$/$s$^2$ directed, approximately, towards the Sun
\cite{LPDSolarSystem,Anderson2002}. Despite the many suggestions for new physics \cite{TuryshevReview}, the problem was finally
settled, after the careful retrieval of the whole telemetry dataset, as originating from the anisotropic emission of heat from
the radioactive sources on the thermoelectric generators \cite{Rievers2011,PioneerPRL,Bertolami2010}.

Even more intriguing is the flyby anomaly, i. e., the unexplained difference among the post-encounter and the pre-encounter Doppler residuals of a spacecraft in a gravity assist manoeuver around the Earth \cite{Anderson2008}. The first detection of the effect occurred during the first Galileo flyby of the Earth on December, 8th, 1990. In this case the discrepancy was interpreted as an anomalous increase of $3.92$ mm$/$s in the post-encounter asymptotic velocity. It is important to emphasize that this anomaly is also observed in the ranging data and cannot be attributed to a conventional or unconventional issue related entirely
to the Doppler tracking. A primary evaluation of the possible conventional physical effects with could be contributing to the anomaly was carried out by  L\"ammerzahl et al. \cite{LPDSolarSystem}. Ocean tides and a coupling of the spacecraft to the
tesseral harmonic terms in the geopotential model have also recently been studied \cite{AcedoMNRAS}. Atmospheric friction can also be dismissed except for flybys at altitudes of $300$ km or lower \cite{Acedo2017one}. The same can be said of the corrections corresponding to General Relativity \cite{IorioSRE2009,Hackmann}, thermal effects \cite{Rievers2011} or other
\cite{Atchison}.

The absence of any convincing explanation have motivated many researchers to undertake the task of looking for models
beyond standard physics. An early work by  Adler \cite{Adler2010,Adler2011} presented a model in which a halo of
dark matter coalesces around the Earth and its interactions would explain away the flyby anomaly.
Anaway, these interactions would verify very stringent conditions. We have also many models which refer to
extensions of General Relativity or modifications of standard newtonian gravity: extensions of Whitehead's theory
of gravity \cite{Acedo2015,Acedo2017three}, topological torsion \cite{Pinheiro2014,Pinheiro2016}, retardation effects \cite{Hafele}, motion in conformal gravity \cite{Varieschi2014} or some {\em ad hoc} modifications of the Newtonian potential
\cite{Nyambuya2008,Wilhelm2015,Bertolami2016}. In the work of Bertolami et al. \cite{Bertolami2016} several ungravity inspired modifications of the Newtonian potential through couplings of the stress-energy tensor or the baryonic current with a rank-$2$ tensor are considered. However, the authors conclude that no modifications of the classical Newtonian potential of this kind can account for the anomalous energy changes detected during the flybys. Consequently, dissipative or velocity-dependent effects accounting for an energy transfer from the spacecraft to the planet should be considered in future studies if the anomalies persist after rigorous analysis. One of the objectives of the present paper is to develop a method from which, in principle, we can infer the form of the perturbation from the trajectory. This way we can test if the perturbation is compatible with a conservative force of takes another form as proposed by Bertolami et al. \cite{Bertolami2016} and other authors \cite{Acedo2015}.

Another non-standard model has been developed by McCulloch who considers a modified inertia as a consequence of a Hubble Casimir effect (MiHsC model). This model predicts a qualitative agreement with the anomalous velocity change found in some missions \cite{McCulloch} and it has also been applied to the problem of the rotation of galaxies to predict the velocity curve profile in 
the absence of any dark matter \cite{McCulloch2017}.

This top-down approach from new theoretical models to fit the data for the
anomaly is unlikely to be successful at the present state of research in this area. Although the observations of the anomaly
are clear in some cases, it is still on the threshold of detectability (or it is simply absent) from other flyby manoeuvers
(such as the Juno flyby of the Earth on October, 2013 \cite{Jouannic,Thompson}). It seems more reasonable to improve the
analysis of the flyby trajectories performed around the Earth and to carry out more analysis of other flyby manoeuvers in the
future. This would help to clarify the existence of such an anomaly, its relation to standard gravity and its manifestation
in missions to other planets. The very nature of this anomaly, with its variations in sign and magnitude from flyby to flyby, has 
made very difficult to find a consistent pattern among them \cite{Anderson2008} in order to settle its characteristics and
phenomenology.                                                                                                                                                                                             

This could have been done by a dedicated science mission such as the, now cancelled, Space-Time Explorer and Quantum Equivalence Principle Space Test (STE-QUEST) spacecraft \cite{STEQUEST}. But, as gravity assist manoeuvers are almost routine in every interplanetary mission, we can expect that the necessary data to establish the undeniable existence of the phenomenon and its
anomalous nature, i.e., the lack of explanation within the current paradigm of physics. To achieve this objective, it would
be highly useful to find that similar anomalies are found in the flybys of other planets. If these anomalies are revealed in 
this situation, and as L\"ammerzahl et al. have already claimed \cite{LPDSolarSystem}, we will have an important science case.
Nowadays, the Juno spacecraft is orbiting around Jupiter in a highly elliptical orbit with perido $53.5$ days after the successful orbit insertion on past July, 4th, 2016. After a failed period reduction manoeuver in its second perijove, the spacecraft is now planned to complete a total of $12$ orbits of which six have now been completed. The interesting fact, in connection with out problem, is that Juno is achieving its periapsis at only $4200$ km over the planet top clouds \cite{JunoMissionI,JunoMissionII,JunoMissionIII} and it provides a new opportunity to test the accuracy of orbit determination and the presence of unexpected discrepancies.

One of the problems with the analysis of the flyby anomaly is the scarcity of the data and the absence of dedicated missions to study this phenomenological issue. On the other hand, this does not prevent us from defining a clear-cut research objective in
experimental gravity and space research: Are highly elliptical and hyperbolic orbits with periapsis close to the main body well 
described by our current theories of gravity and spacecraft navigation models ?. Starting with Anderson et al. \cite{Anderson2008} there are many researchers who think that we face a problem in this case and that further research is necessary to obtain as accurate predictions as our current technology allows.

The objective of this paper is to develop an orbital model specially suited for the perijove time-frame. This model should take into account, at least, the tidal effects of Jupiter's Galilean satellites and the known zonal harmonics of the planet. By comparing with the telemetry data we disclose a small, but significant, anomalous acceleration whose components in spherical coordinates are of the order of magnitude of a few mm$/$s$^2$ and decay below the measurement error bars after a period of $30$ minutes before or after the perijove. As we will see this is compatible with the expected order of magnitude from Anderson's phenomenological formula \cite{Anderson2008} and some modified models of gravity \cite{Acedo2015,Acedo2017three}.

\section{Orbital model}
\label{sec:2}

In this section we discuss the development of an orbital model optimized for the region around the perijove. Our problem
is summarized in the set of Newtonian equations of motion:
\begin{eqnarray}
\label{eqmotionr}
\displaystyle\frac{d {\bf r}}{d t}&=&{\bf v} \; ,\\
\noalign{\smallskip}
\label{eqmotionv}
\displaystyle\frac{d {\bf v}}{d t}&=&-\mu_J \displaystyle\frac{{\bf r}}{r^3}+\bm{\mathcal{F}}_{\mbox{tidal}}+\bm{\mathcal{F}}_{\mbox{zonal}} \; ,
\end{eqnarray}
where ${\bf r}$ and ${\bf v}$ are the position and velocity vectors of the spacecraft, respectively, $\mu_J$ is Jupiter's mass
constant, ${\bm{\mathcal{F}}}_{\mbox{tidal}}$ is the perturbing tidal force exerted by the Sun and Jupiter's satellites, and 
$\bm{\mathcal{F}}_{\mbox{zonal}}$ are the corrections arising from the known zonal harmonics of the planet.
Concerning the mass constants of Jupiter, the Sun and the Galilean satellites we have the following values \cite{DE431,Satellites}:
\begin{equation}
\begin{array}{rcl}
\mu_J&=&126712764.800000 \; \mbox{km$^3/$s$^2$}\; , \\
\noalign{\smallskip}
\mu_{\mbox{Sun}}&=&132712440041.939400\; \mbox{km$^3/$s$^2$}\; , \\
\noalign{\smallskip}
\mu_{\mbox{Io}}&=&5959.916\; \mbox{km$^3/$s$^2$}\; , \\
\noalign{\smallskip}
\mu_{\mbox{Europa}}&=&3202.739\; \mbox{km$^3/$s$^2$}\; , \\
\noalign{\smallskip}
\mu_{\mbox{Callisto}}&=&7179.289\; \mbox{km$^3/$s$^2$}\; , \\
\noalign{\smallskip}
\mu_{\mbox{Ganymede}}&=&9887.834\; \mbox{km$^3/$s$^2$}\; . 
\end{array}
\end{equation}
The tidal force on the reference frame of Jupiter exerted by the Sun or any of Jupiter's satellites is given by:
\begin{equation}
\label{Ftid}
\bm{\mathcal{F}}_{\mbox{tidal}}=\mu \left(-\displaystyle\frac{{\bf R}}{R^3}+\displaystyle\frac{{\bf R}-{\bf r}}{\left(r^2+R^2-2 {\bf r} 
\cdot {\bf R} \right)^{3/2}}\right)\; .
\end{equation}
Here ${\bf R}$ is the position vector of the third body and ${\bf r}$ is the position vector of the spacecraft with respect to
the mass center of Jupiter. The contribution to the gravitational potential of the quadrupole, octupole and higher order terms
is given by:
\begin{equation}
\label{Upot}
U(r,\theta)=-\displaystyle\frac{\mu_J}{r} \, \displaystyle\sum_{n=2}^N \, J_n \, \left( \displaystyle\frac{R_J}{r} \right)^n P_{n}(\cos \theta) \; ,
\end{equation}
where $J_n$ are the zonal harmonics coefficients \cite{Vallado}, $P_n(x)$ are the Legendre polynomials and $\theta$ is the
colatitude of the spacecraft (the angle formed by the spacecraft's position vector and the axis of the planet). The reference
radius is $R_J=71492$ km and the known zonal harmonics \cite{Transfer} are given by:
\begin{equation}
\label{Jcoeff}
\begin{array}{rcl}
J_2&=&0.01469645 \; , \\
\noalign{\smallskip}
J_4&=&-0.00058722\; , \\
\noalign{\smallskip}
J_6&=&0.00003508\; ,
\end{array}
\end{equation}
so, we take $N=6$ in Eq. (\ref{Upot}) and we consider also only the coefficients of even order. In spherical coordinates, the components of the perturbing force corresponding to the potential in Eq. (\ref{Upot}) are:
\begin{eqnarray}
\label{Fgeor}
{\mathcal F}_r &=&-\displaystyle\frac{\mu_J}{r^2} \, \displaystyle\sum_{n=2}^N \,
J_n (n+1)\left( \displaystyle\frac{R_J}{r} \right)^n \, P_n(\cos \theta) \; ,\\
\noalign{\smallskip}
\label{Fgeot}
{\mathcal F}_\theta &=&-\displaystyle\frac{\mu_J}{r^2}\, \displaystyle\sum_{n=2}^N  \, J_n \left(\displaystyle\frac{R_J}{r} \right)^n\, P^{'}_n(\cos \theta) \, \sin\theta\, \; .
\end{eqnarray}
So the perturbing force arising from the zonal harmonics terms is $\bm{\mathcal{F}}_{\mbox{geo}}=
\mathcal{F}_r \, \hat{\bm{r}}+\mathcal{F}_\theta \, \hat{\bm{\theta}}$, with $\hat{\bm{r}}$ and $\hat{\bm{\theta}}$ as the
unit radial and polar vectors. 
In order to calculate this force we must know the orientation of the axis of Jupiter in the ecliptic frame of reference. 
The right ascension, $\alpha_J$, and declination, $\delta_J$, of the unit vector pointing in the direction of this axis vary with time as a consequence
of precession and nutation and it is given by \cite{Jupiterfacts}:
\begin{eqnarray}
\label{EqaxisJ}
\alpha_J &=& 268.057 - 0.006\, T\; ,\\
\noalign{\smallskip}
\delta_J &=& 64.495 + 0.002\, T\; ,
\end{eqnarray}
where $T$ is the time in Julian years from the J2000 reference date (Julian day $2451545.0$). For the first flyby of Jupiter on
August 27th, 2016 we have $T=0.0166516$ Julian years and this allows for a determination of the axis orientation in the Earth's equatorial frame of reference with an accuracy of $0.001$ sexagesimal degrees. 
The obliquity of the ecliptic is also known with high accuracy at a given Julian date in terms of the following polynomial in $T$ \cite{Almanac}:
\begin{equation}
\label{oblq}
\begin{array}{rcl}
\chi&=& 23^\circ 26^{'} 21.406^{''}-46.836769^{''} T \\
\noalign{\smallskip}
&-&0.0001831^{''} T^2+0.00200340^{''} T^3 \\
\noalign{\smallskip}
&-&5.76^{''}\times 10^{-7} T^4-4.34^{''} \times  10^{-8} T^5\; .
\end{array}
\end{equation}
From Eqs. (\ref{EqaxisJ}) and (\ref{oblq}) we have calculated the components of Jupiter's axis in the ecliptic reference frame:
\begin{equation}
\label{kaxis}
\begin{array}{rcl}
\hat{k}_x&=&-0.01460\; ,\\
\noalign{\smallskip}
\hat{k}_y&=&-0.03582\; ,\\
\noalign{\smallskip}
\hat{k}_z&=&0.99925\; .
\end{array}
\end{equation}
Finally, from the vector in Eq. (\ref{kaxis}) and the spacecraft's position vector we can determine the colatitude angle in Eqs. (\ref{Fgeor}) and (\ref{Fgeot}) in order to compute the force arising from the zonal harmonics.

We will also point out that we have used an iterative procedure to solve the equations of motion in Eq. (\ref{eqmotionr})-(\ref{eqmotionv}) because its simplicity and stability in comparison with the alternative approach in which the unknown functions, ${\bf r}(t)$ and ${\bf v}(t)$, appear also in the perturbation terms. The algorithm proceeds as follows:
\begin{itemize}
\item We select a given timestep (in minutes in the ephemeris for Juno \cite{Horizons}) as close to the perijove as possible.
This would be our initial condition. The fact that it does not coincide with the perijove is not relevant for our purpose.

The main reason for this backwards and forward integration procedure is to
reduce the propagation of errors in the numerical method. Alternatively, we can start from an instant $180$ minutes before the 
perijove and integrate throughout the perijove to another instant $180$ minutes after the perijove but we have found that numerical errors are larger in this second method.

\item The equations of motion are integrated backwards and forward in time for a period of, at least, $180$ minutes. In this
first integration we ignore the perturbation forces.
\item The tidal forces and the zonal contribution to the perturbation are evaluated at the positions given by the zeroth-order keplerian approximation (the ideal hyperbolic orbit).
\item A new integration of the equations of motion is carried out with the perturbing forces evaluated in the previous step. This
would be our first order approximation.
\item Subsequently, we evaluate the perturbation forces with the positions obtained in the $n$th-order approximation to obtain
the position and velocities of the spacecraft in the $(n+1)$th-order approximation. 
\item The algorithm stops when the differences among the $n$th-order and the $n+1$th-order approximation is below a given threshold.
\end{itemize}
We must also emphasize that computation was carried out with double precision to keep the accuracy of the model data throughout the evaluation of the predictions of the model. In the next section we will discuss the results obtained with the methods summarized here.

\section{Evaluation of the residual acceleration at the perijove}
\label{sec:3}

In this section we will discuss the analysis of the first, third and fourth orbits of Juno around Jupiter. Using the
method described in the previous section we will focus on the region around the perijove in order to 
unveil any possible anomalies in the trajectory as they have already been found in close flybys of the Earth \cite{Anderson2008}.
We have not considered the second flyby in which a period reduction manoeuver was planned but, later on, cancelled because
the helium check valves were not operating properly \cite{Junovalve}. Consequently, the spacecraft was set into safe mode during that particular flyby.

We considered the telemetry data for the first flyby starting from August, 26th at $00\mbox{:}00.000$ Barycentric dynamical time (TDB). From such reference the minimum distance to the center of Jupiter was attained at minute $t_P=2212$. The spatial coordinates and velocity at that instant were taken as the initial conditions for our integration procedure. 

\begin{figure}
\includegraphics[width=\columnwidth]{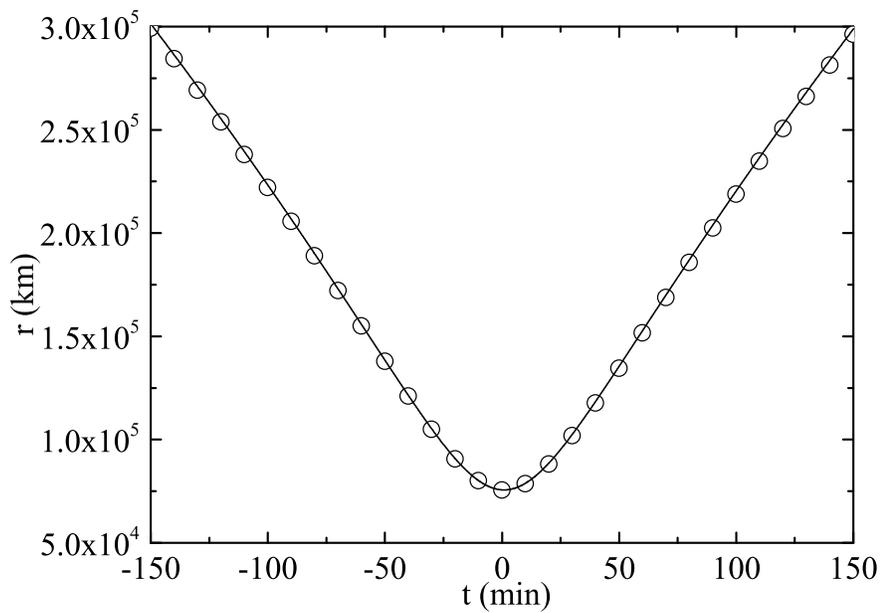}
\caption{Comparison among the distance to the center of Jupiter of the Juno spacecraft during the first flyby (open circles) and the ideal hyperbolic approximation (solid line). Coordinate $r$ is measured in km and time in minutes since the initial condition close to the perijove.}
\label{fig1}
\end{figure}

If we ignore the perturbation terms in Eqs. (\ref{eqmotionr})-(\ref{eqmotionv}) we obtain the ideal hyperbolic keplerian solution
as a crude approximation to the real trajectory. As shown in Fig. \ref{fig1} the difference seems small, in the distance 
scale of Jupiter's radius, but it is critical in our analysis of the trajectories.

\subsection{Tidal forces}
\begin{figure}
	\includegraphics[width=\columnwidth]{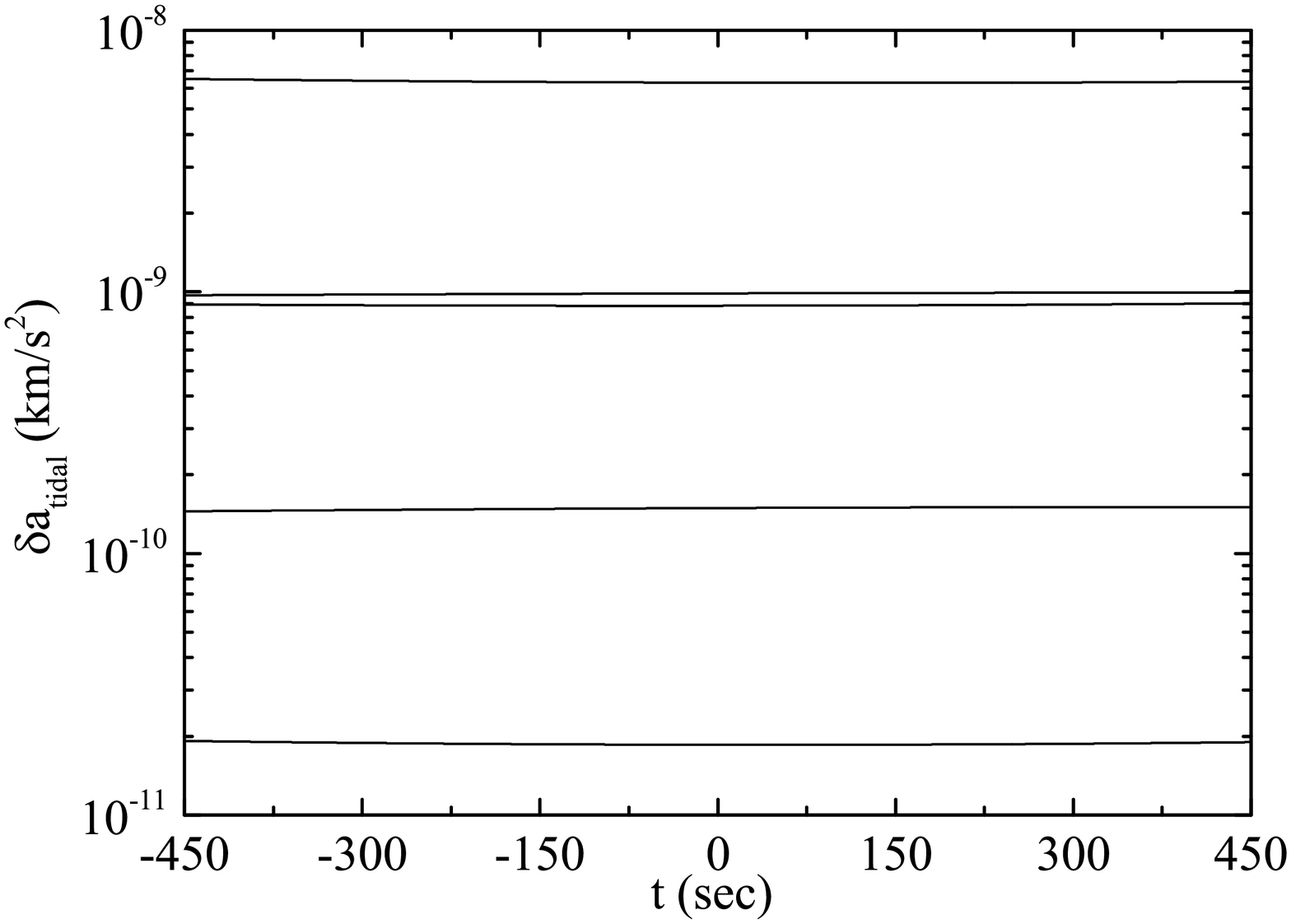}
   \caption{Tidal acceleration exerted upon the Juno spacecraft by (from top to bottom): Io, Ganymede, Europa, Callisto and the Sun. The tidal force per unit mass is measured in km$/$s$^2$ and time is given in seconds from the perijove.  
}
\label{fig2}
\end{figure}

It is convenient to consider separately the effect of tidal forces to compare its impact on the trajectory perturbations with
that of the zonal harmonics. We will see that in the vicinity of the perijove the tidal contribution is small in relation to
the effect of the multipole terms in the gravitational model of Jupiter. To visualize the magnitude of the different tidal
forces we have plotted in Fig. \ref{fig2} the magnitude of the tidal forces exerted by any of the Galilean satellites and, also, by the Sun.
 
Notice that the distance to the planet and the spacecraft during the flyby seems the most important factor on the determination
of the magnitude of these forces. At an average distance of $5.2$ Astronomical Units from the Sun, the tidal effect is less important than in the case of Earth flybys \cite{Anderson2008}. As Io is the closest satellite it also gives the larger tides, despite it is not as massive as Ganymede or Callisto.

\begin{figure}
	\includegraphics[width=\columnwidth]{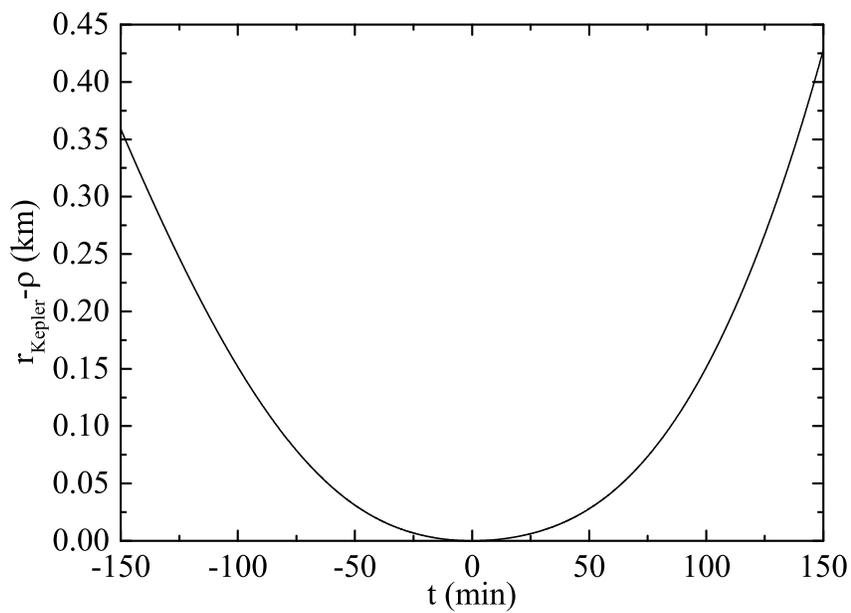}
   \caption{The difference among the distance of the Juno spacecraft to the center of Jupiter in the ideal hyperbolic approximation and the prediction of the orbital model (including only the effect of the tidal forces). This difference is measured in km and time in minutes since the perijove.}
   \label{fig3}
\end{figure}

In Fig. \ref{fig3} we have plotted the results for the orbital model incorporating only the effect of the tides from Io, Europa, Ganymede, Callisto and the Sun. We show that there is a reduction, in the correct direction according to Fig. \ref{fig1}, for the prediction of the radial coordinate of the spacecraft. However, this is not sufficient to provide a
good fit of the discrepancies. 
\subsection{Results for the complete orbital model}

\begin{figure}
	\includegraphics[width=\columnwidth]{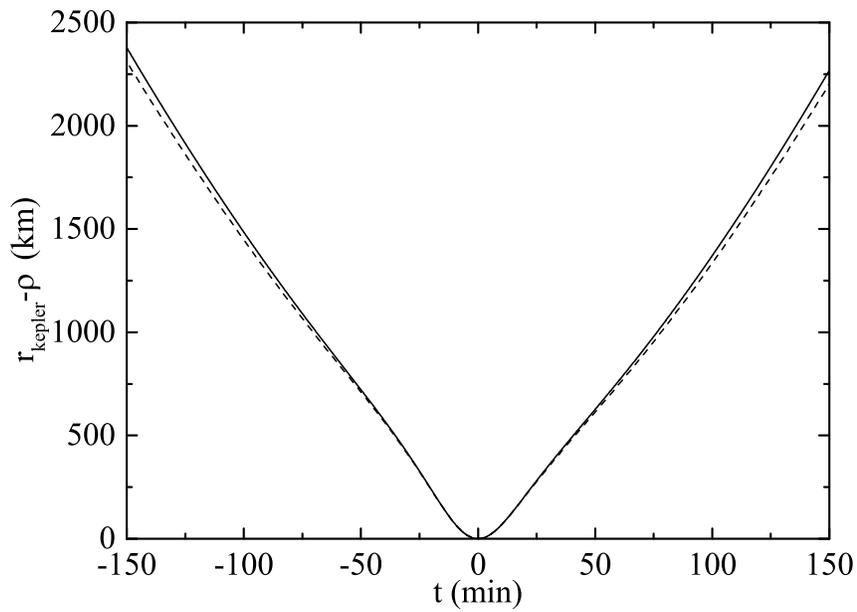}
   \caption{Difference among the radial coordinate (in the ideal keplerian approximation) and the data (solid line) compared with the same substraction evaluated for the orbital model (dashed line). Notice that the agreement among the data and the model is good except for small, but noticeable discrepancies, that build up before or after the perijove.}
   \label{fig4}
\end{figure}

\begin{figure}
	\includegraphics[width=\columnwidth]{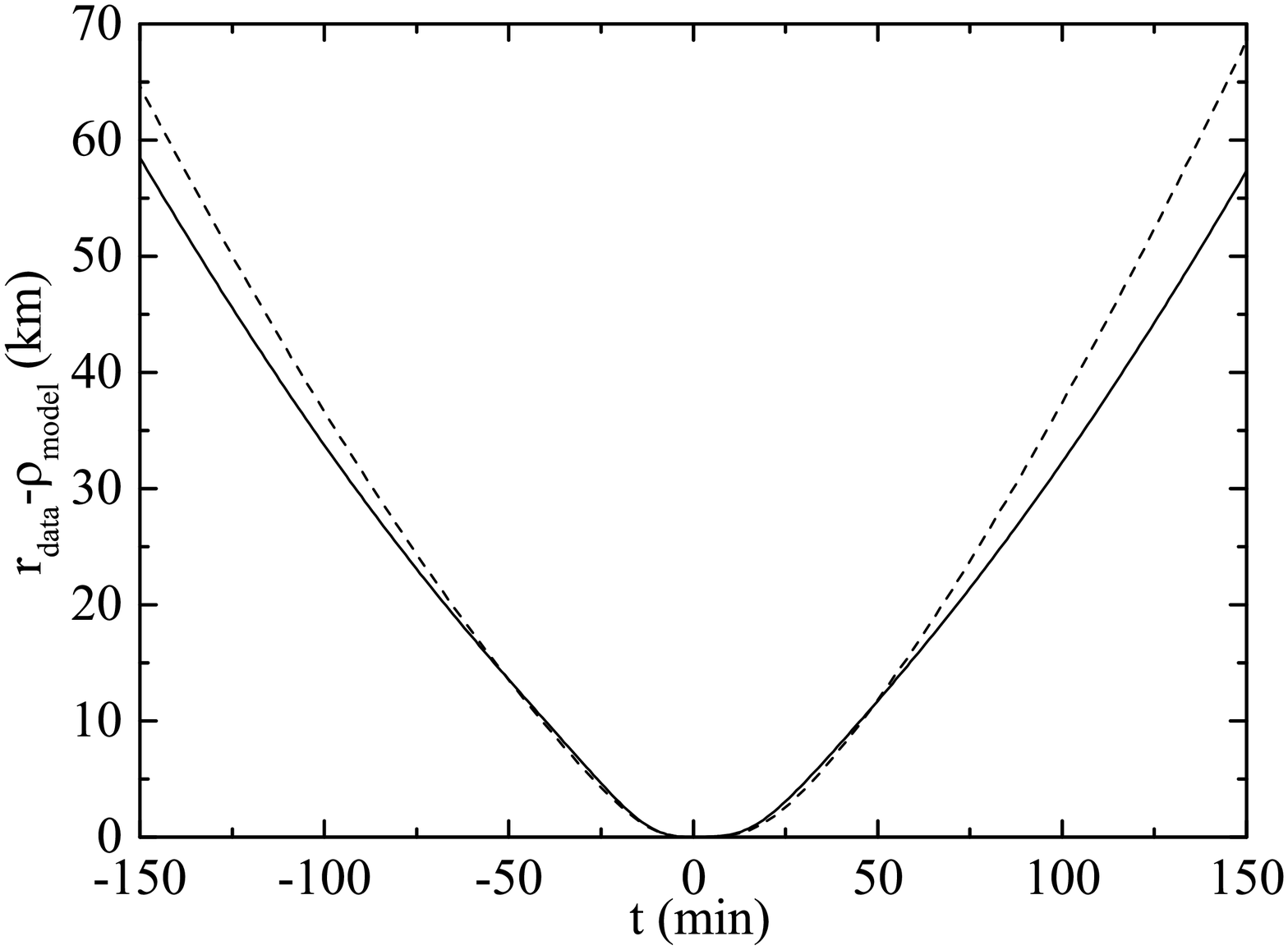}
   \caption{Difference among the data for the radial coordinate and the orbital model: first approximation (dashed line) and
   second approximation (solid line). Distances are measured in km and time in minutes from the perijove.}
   \label{fig5}
\end{figure}

Our interest is now to implement the whole orbital model as defined in Sec. \ref{sec:2}. In the first place, we have plotted the
difference among the radial coordinate in the zeroth-order keplerian approximation and the data compared with the same difference for the prediction of the orbital model.  This is shown in Fig. \ref{fig4}. We see that the agreement is very good but the model
systematically underestimates the altitude of the spacecraft as if some outwards anomalous radial acceleration were acting upon Juno during its approximation to the perijove.
This result requires further investigation so we have shown in Fig. \ref{fig5} the difference of the model predictions directly with the data for the first and second iteration of the algorithm described in Sec. \ref{sec:2}. As we will see later, the third and subsequent iterations yield only very small corrections to this picture.
The discrepancy is around several tens of kms and, consequently, it can be considered a very clear signal in the telemetry data that deserves further analysis. The difference among the model and the data is sufficiently accurate to allow for a determination
of the components of the acceleration field responsible for this deviation of the spacecraft from the predicted trajectory.
This is achieved by using a fourth-order central finite difference method \cite{Fornberg}:
\begin{equation}
\label{deltaa}
\begin{array}{rcl}
\delta{\bf a}&=&\displaystyle\frac{1}{h^2} \,\left\{ -\displaystyle\frac{1}{12}\left( \delta {\bf r}(t - 2 h)+
\delta {\bf r}(t+2 h)\right)\right. \\
\noalign{\smallskip}
 &+&\displaystyle\frac{4}{3}\left( \delta {\bf r}(t-h)+\delta {\bf r}(t+h) \right)-
\left. \displaystyle\frac{5}{2} \delta {\bf r}(t)\right\} \\
\noalign{\smallskip}
&-&\displaystyle\frac{1}{90} \displaystyle\frac{d^6 \delta {\bf r}}{d t^6} \, h^4+ {\mathcal O}\left(h^5\right) \; ,
\end{array}
\end{equation}
where $h$ is the timestep and the error term can be estimated by using the corresponding approximation for the sixth-order
derivative.  As the data provided for the spacecraft tracking is separated by one minute intervals \cite{Horizons} we should also
choose $h=1$ min to allow for the evaluation of Eq. (\ref{deltaa}) with the same accuracy. 
\begin{figure}
	\includegraphics[width=\columnwidth]{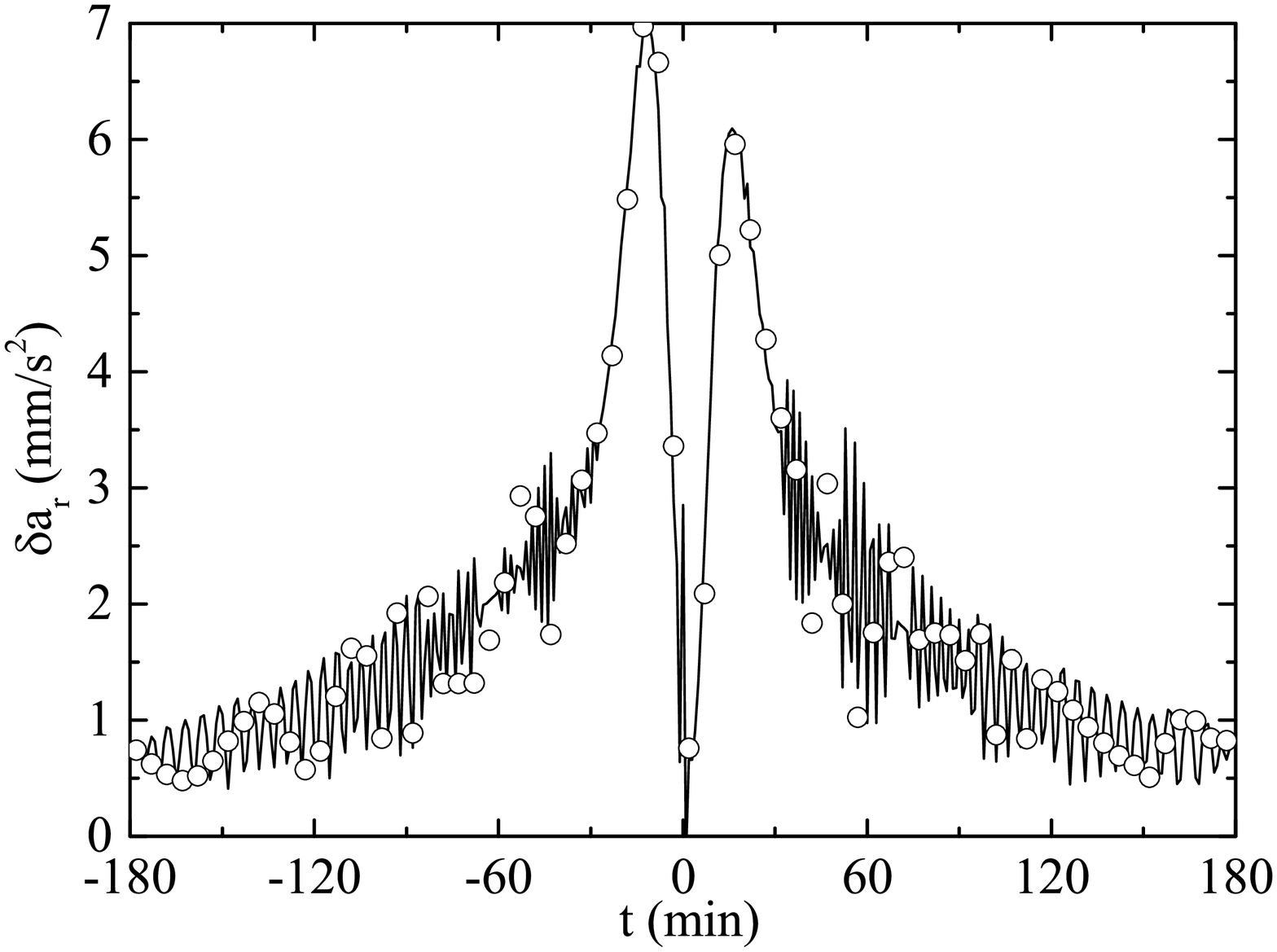}
   \caption{Radial component of the anomalous acceleration acting upon the Juno spacecraft during the perijove manoeuvre. The solid line corresponds to the first flyby and the open circles to the third one.}
   \label{fig6}
\end{figure}

\begin{figure}
	\includegraphics[width=\columnwidth]{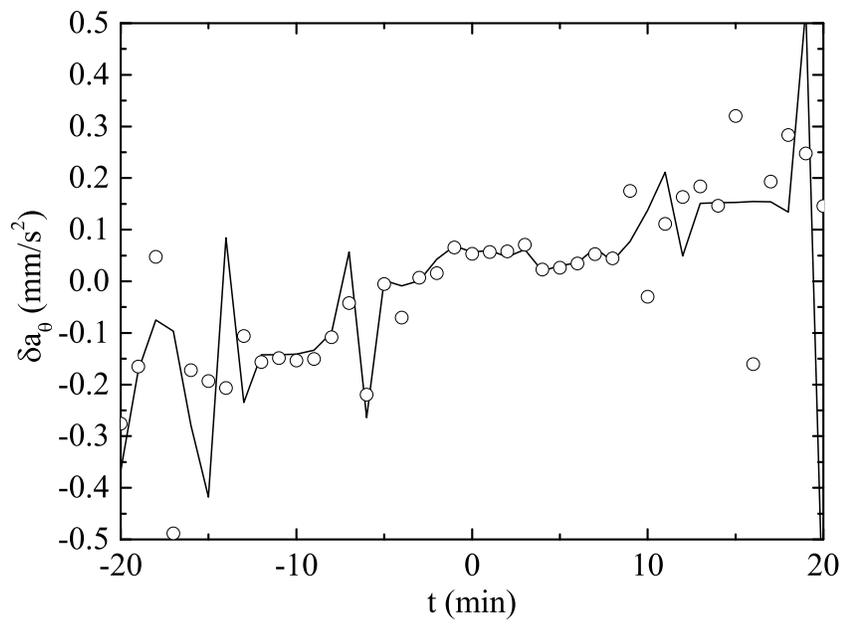}
   \caption{The same as Fig. \protect\ref{fig6} but for the polar component.}
   \label{fig7}
\end{figure}

\begin{figure}
	\includegraphics[width=\columnwidth]{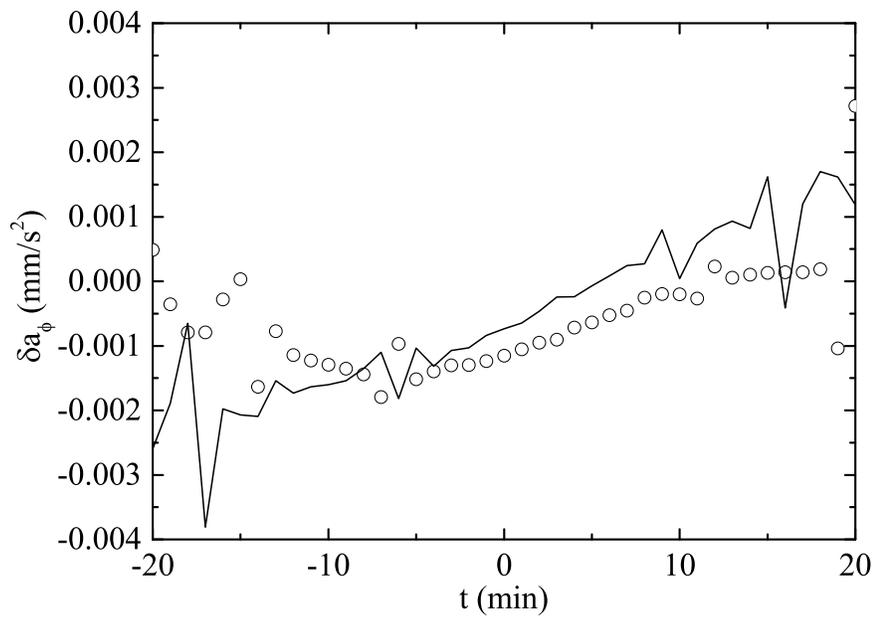}
   \caption{The same as Figs. \protect\ref{fig6} and Fig. \protect\ref{fig7} but for the azimuthal component.}
   \label{fig8}
\end{figure}

By computing the extra acceleration in Eq. (\ref{deltaa}) to match the two trajectories (the one in our orbital model and the one delivered by the Juno's mission team to JPL fitted to the telemetry data) we find the discrepancies shown in Figs. \ref{fig6}-\ref{fig8} for the radial, polar and azimuthal components. In these figures we show the results for the first and third flyby of the Juno spacecraft which, essentially, followed the same trajectory in the two approximations to Jupiter. Notice that the results are similar for both flybys. In the case of the radial component, we find two sharp peaks of different amplitude as a manifestation of an oscillatory behaviour as a function of time. On the other hand, the analysis yields different behaviour for the polar and azimuthal components as
shown in Fig. \ref{fig7} and Fig. \ref{fig8} but these are two or three orders of magnitude smaller than the radial component and we can think that this is not statistically significant as other sources of error may also be distorting these components.

\subsection{Sources of error and interpretation of the results}

In this section we will discuss some possible sources of error that could explain the discrepancy among the trajectory fitted by 
the Juno's team and the orbital model proposed in this paper. And, in particular, the perturbing acceleration whose components
in spherical coordinates are plotted in Figs. \ref{fig6}-\ref{fig8}. 
\begin{figure}
\includegraphics[width=\columnwidth]{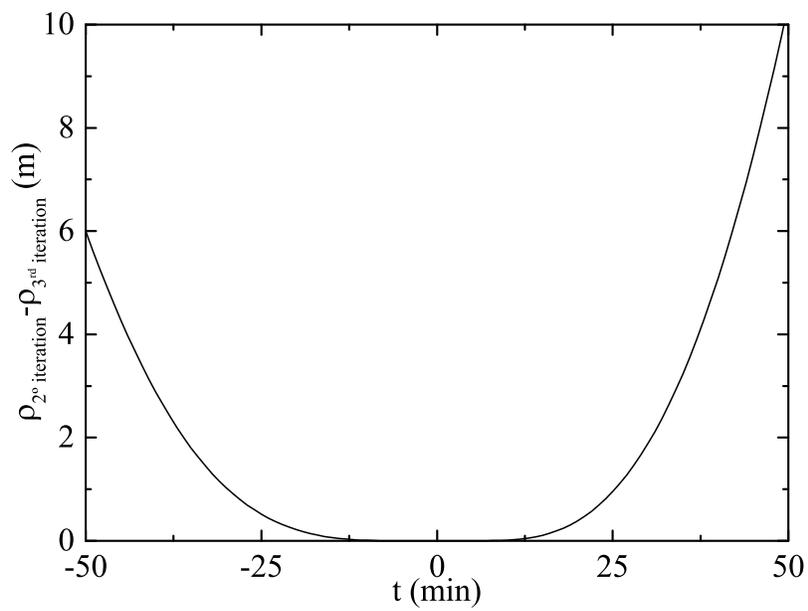}
   \caption{The difference (in meters) between the radial coordinate for the second and the third iteration of the Picard's method discussed in the text. Time is measured in minutes from the closes approach to Jupiter in the first Juno's flyby.}
   \label{fig9}
\end{figure}
In Fig. \ref{fig9} we have shown the difference among the second and the third iteration of the Picard's method applied to the
solution of the equations of motion Eq. (\ref{eqmotionr})-(\ref{eqmotionv}) that, for a period of fifty minutes around the perigee, is of the order of $10$ meters and, consequently, three orders of magnitude below the differences of the second iteration and the JPL's fitting as shown in Fig. \ref{fig5}. So, we can be confident that the convergence of the method
is very fast for our problem and the results are reliable.

Another question is the accuracy of the ephemeris of the moons of Jupiter. Since the beginning of radar astronomy the precision of these measurements has improved very fast. As early as 1965 the features of Venus were tracked with a maximum uncertainty of
$3$ km \cite{Tausworthe} by Deep-Space radars. Subsequent missions to Jupiter has allowed also a high-accuracy determination of
the orbits of the moons of this planet. Starting from Eq. (\ref{Ftid}) we get an estimation of the perturbation of the tidal forces as a function of the uncertainty in the position $\delta {\bf R}$:

\begin{equation}
\begin{array}{rcl}
\delta \bm{\mathcal{F}}_{\mbox{tidal}}&=&\mu \left(-\displaystyle\frac{\delta {\bf R}}{R^3}+3 \displaystyle\frac{{\bf R}}{R^5}{\bf R} \cdot \delta {\bf R}\right.\\
\noalign{\smallskip}
&+&\displaystyle\frac{\delta {\bf R}}{\left(r^2+R^2-2 {\bf r}
\cdot {\bf R} \right)^{3/2}} \\
\noalign{\smallskip}
&-&\left. 3 \displaystyle\frac{\left({\bf R}-{\bf r}\right) \cdot \delta {\bf R}}{\left(r^2+R^2-2 {\bf r}
\cdot {\bf R} \right)^{5/2}} \left( {\bf R}-{\bf r}\right)   \right)\; .
\end{array}
\end{equation}

Assuming an error of $3$ km in a random direction for the position of Io we get an uncertainty in the tidal acceleration exerted upon Juno of $10^{-13}$ km$/$s$^2$, which it is certainly very small and can be dismissed as the origin of the possible anomaly
discussed in this paper.

Mismodelling of the zonal coefficients is also a source of error. For example, a variation of $J_2$ by $10^{-8}$ would
imply, according to Eqs. (\ref{Fgeor}) and (\ref{Fgeot}), a perturbing force of magnitude $\vert \delta {\bf F} \vert
\simeq 2.92 \times 10^{-4}$ mm$/$s$^2$, which it is very small in comparison perturbing accelerations we have found in the previous section. However, a disregarded zonal coefficient with order $J_n \simeq 10^{-4}$ could explain the anomalies in
the integration of our model. So, further research into the structure of Jupiter is  necessary and this can be achieved in
future analysis of the data provided by the Juno mission. Anyway, it seems unlikely that zonal coefficients of order eight and
higher could be so large. As it happens in the case of the Earth, we expect that these coefficients would diminish with the order and from Eq. (\ref{Jcoeff}) an upper bound $J_8 < 10^{-5}$ seems reasonable for the first ignored coefficient in our calculation.
The Eqs. (\ref{Fgeor}) and (\ref{Fgeot}) then gives us an estimation of $1$ mm/s$^2$ for the magnitude of the component of the
acceleration at perijove but this is only $10^{-3}$ mm/s$^2$ an hour before or after the perijove. So, a better modelling of the
gravitational model of Jupiter is necessary for studying the orbit of Juno near the perigee but we cannot discard the anomaly
because it persists even an hour after crossing the perigee as shown in Fig. \ref{fig6} with a magnitude too large to be explained only in terms of mismodelled or ignored zonal coefficients.

Another source of mismodelling can arise from the estimation of Jupiter's axis orientation in space. If we consider that the
axis at J2000 instead of the correction for the date of Juno's flybys, by taking $T=0$ in Eq. (\ref{EqaxisJ}), a perturbation
in the force term of the potential model of magnitude $\vert \delta {\bf F} \vert \simeq 5.63 \times 10^{-5}$ mm$/$s$^2$ is found.

One should also consider that the spacecraft is an extended object which rotates at three revolutions per minute. This would 
generate a small magnetic moment for Juno which could contribute to the equations of motion through interaction with the magnetic
field of Jupiter but this has been estimated as negligible in other cases \cite{LPDSolarSystem}. It has also been shown that 
helicity of radio waves can exhibit a coupling with the rotation of the spacecraft and the rotation of the planet \cite{Helicity} but this only influences the two-way Doppler data and it can not explain the arising of the anomaly also in the
ranging data \cite{LPDSolarSystem}.

A remaining possibility is the connection among the discrepancies found and the flyby anomaly, which have been detected earlier
in spacecraft flybys of the Earth. Some models have suggested that the Earth's gravitational field is distorted by an
unknown extra term, not taken into account in General Relativity, and that this can be interpreted as a force field with a
range of a few hundred kms \cite{Acedo2017two}. In the model by Acedo and Bel \cite{Acedo2015,Acedo2017three} an anomalous
azimuthal component of the gravity acceleration is proposed. The magnitude of this extra acceleration is given by:
\begin{equation}
\label{Bel}
\delta a=\displaystyle\frac{\mu}{r^2} \, \displaystyle\frac{\Omega R}{c} \; ,
\end{equation}
where $\mu$, $R$ are the mass constant and radius of the planet, $\Omega$ is the angular velocity with respect to the fixed stars and $r$ the distance of the spacecraft to the center. It was shown that this model yields a qualitative agreement with the
measured anomalies in several flybys of the Earth \cite{Acedo2015}. If we apply this expression to the case of Jupiter, by taking
into account that $r \simeq R = 71492$ km at the perijove and that the Jupiter's angular velocity is $\Omega=2 \pi/T$, $T=9.9259$ hours \cite{Jupiterfacts}, we get $\delta a=1.0396$ mm$/$s$^2$. This agrees with the order of magnitude of the peaks in the radial component of the extra acceleration shown in Fig. \ref{fig6}. If this is merely a numerical coincidence or we require a fundamental modification in our understanding of highly elliptical orbital dynamics could only be disclosed by further 
analysis of these trajectories in future missions.

Finally, some possible classical effects and the magnitude of the acceleration imparted upon the spacecraft are listed in Table \ref{tab1} in order to compare with the anomaly. Some of these values are taken from L\"ammerzahl et al. study for the Earth's flyby anomaly \cite{LPDSolarSystem}, but they can be extrapolated to the case of Juno at Jupiter.

\begin{table}
\centering
\caption{Non-modelled classical effects in our orbital model and the magnitude of the corresponding accelerations.}
\begin{tabular}{lc}
\label{tab1}
{\em Non-modelled effect} & {\em Acceleration's magnitude} \\
\noalign{\smallskip}
Solar wind & $10^{-7}$ mm$/$s$^2$ \\
\noalign{\smallskip}
Albedo's pressure & $10^{-6}$ mm$/$s$^2$ \\
\noalign{\smallskip}
Magnetic moment & $10^{-12}$ mm$/$s$^2$ \\
\noalign{\smallskip}
Spacecraft's charge & $10^{-5}$ mm$/$s$^2$ \\
\noalign{\smallskip}
Atmospheric's friction & $10^{-5}$ mm$/$s$^2$ \\
\noalign{\smallskip}
Tides & $10^{-2}$ mm$/$s$^2$ \\
\noalign{\smallskip}
Ephemeris' uncertainty & $10^{-7}$ mm$/$s$^2$ \\
\noalign{\smallskip}
Ignored zonal harmonics & $1$ mm$/$s$^2$ \\
\noalign{\smallskip}
Gravitomagnetism & $0.1$ mm$/$s$^2$ 
\end{tabular}
\end{table}

At this table we see that most effects contribute only a very small fraction to the putative anomalous acceleration disclosed in this work. Solar wind indeed is only around a factor $1/25$ of the contribution at Earth because it decreases with the square of the
distance to the Sun. Additional zonal harmonics to those considered in our model are certainly an important issue to elucidate
in future research about Jupiter's interior as well as the gravitomagnetic effect \cite{Hackmann,IorioSRE2009,IorioJunoLT} but their contribution to the orbital model cannot explain the trajectory as modelled in this paper. 

\section{Conclusions}

Juno mission to Jupiter is becoming one of the most successful space missions of the XXIst century. It is also the first
time in which a spacecraft performs close flybys of a giant planet to analyze its atmosphere, magnetic field and
gravitational structure \cite{JunoMissionI,JunoMissionII,JunoMissionIII}. The Juno spacecraft is currently in
a highly eccentric elliptical orbit around Jupiter. This orbit is perpendicular to Jupiter's equatorial plane and crosses over
the poles of the planet with a periapsis near to the equator. The altitude of the perijove over Jupiter's top clouds is
around $4200$ km for the first flyby and it is programmed to raise slowly throughout the planned $36$ orbits. A period reduction
manoeuver from the $53.5$ days period orbit to an, approximately, $14$ days period orbit was also scheduled but finally
cancelled because a failure in the opening of the helium check valves \cite{Junovalve}.
 
The orbit is also being carefully monitored by the retrieval of telemetry data and the evaluation of the ephemeris from the
mission team, these are then incorporated into the Horizons' web system \cite{Horizons} which make them available to the
whole scientific community. The resulting ephemeris are fits to radiometric tracking data which take into account all the
modelling details taken into account by the navigation team of the particular mission including atmospheric friction, solar pressure and perturbations by the planets and satellites. Our objective in this paper has been to develop an independent 
orbital model for Juno's trajectory in the vicinity of the perijove in order to compare with the orbit fitted by the Juno mission 
team to the telemetry data. In doing so, we should be able to disclose any possible discrepancies and to test the validity
of orbital determination programs.

In our model we have taken into account the tidal forces exerted by the Sun and by Jupiter's larger satellites, i. e., the Galilean satellites:
Io, Europa, Ganymede and Callisto and also the contributions of the known zonal harmonics \cite{Transfer}. We have found that the
multipolar field contributions due to Jupiter's oblateness are far more important near the perijove than the tidal forces and 
that they provide a very good fit of the trajectory. Nevertheless the agreement is not perfect within the error bars for the
models and small discrepancies persist after considering the aforementioned perturbations. We have interpreted this discrepancy
as an anomalous extra acceleration whose component is mainly radial. This acceleration is in the range of a few mm$/$s$^2$ and exhibit two, almost symmetric, peaks around fifteen minutes before and after the perijove. In a period of $3$ hours after the
crossing of the perigee it has decayed near to zero. At this moment, the spacecraft is located at a distance of $\simeq 4.75$ Jupiter's radii. All this made the resulting anomaly consistent with an interaction which decays very fast with the distance to
the planet as it have been suggested in connection with the flyby anomaly \cite{Acedo2017two}. On the other hand, there are
other possible sources which require further investigation such as the mismodelling of zonal coefficients for the planet or
the effect of its strong magnetic field. Anyway, in the case of the magnetic forces they should be directed mainly perpendicular to Juno's trajectory as it flybys the planet in a polar orbit and, on the contrary, the anomaly is found mainly as a radial
component of the acceleration.

Summarizing, we can say that in this paper: (i) We have found evidence that an anomaly could be operating also during the Juno flybys of Jupiter (ii) We have developed a theoretical model to compare with the orbital model fitted to telemetry data in order to disclose the form of the possible anomalous acceleration field acting upon the spacecraft. A significant radial component was found and this decays with the distance to the center of Jupiter as expected from an unknown physical interaction. (iii) The anomaly shows an asymmetry among the incoming and outgoing branches of the trajectory and this could be suggestive of a non-conservative interaction. The confirmation of these conclusions would require further independent analysis and we hope that our work will stimulate future research in this and other planetary flybys.

In the context of this discussion we should also mention that similar anomalous accelerations are also found in several spacecraft flybys of the Earth \cite{AcedoEarth}. In this case, they are only a $1.5$ \% of those found in the case of
the Juno's flybys of Jupiter. This contributes to the interest of the problem of high-accuracy orbital dynamics in the
particular case of close flybys of the planets. Only the interplay among physics, spacecraft navigation and engineering
could finally solve this issue.

\section*{Acknowledgements}

NASA's Jet Propulsion Laboratory and the Juno mission team are ackowledged for providing all the ephemerides of this work through the on-line Horizon system. 

\bibliographystyle{plain}     
\bibliography{acedobiblio}

\end{document}